\documentclass[prd,nofootinbib,twocolumn,amssymb,amsmath]{revtex4}
\usepackage{graphicx} 

\def\mayakranc#1{\textsc{MayaKranc}#1}

\def\newacronym#1#2#3{\gdef#1{#3 (#2)\gdef#1{#2}}}

\newacronym{\NSF}{NSF}{National Science Foundation}
\newacronym{\NASA}{NASA}{National Aeronautics and Space Administration}
\newacronym{\lisa}{LISA}{the Laser Interferometer Space Antenna}
\newacronym{\ligo}{LIGO}{Laser Interferometer Gravitational-wave Observatory} 
\newacronym{\Caltech}{Caltech}{California Institute of Technology}
\newacronym{\MIT}{MIT}{Massachusetts Institute of Technology}
\newacronym{\sph}{SPH}{smooth particle hydrodynamics}
\newacronym{\tsi}{TSI}{the Terascale Supernova Initiative}
\newacronym{\wmap}{WMAP}{the Wilkinson Microwave Anisotropy Probe}
\newacronym{\decigo}{DECIGO}{the Deci-Hertz Interferometric Gravitational-wave Observatory} 
\newacronym{\cmbr}{CMBR}{cosmic microwave background}
\newacronym{\ibbh}{IBBH}{intermediate binary black hole}
\newacronym{\bdj}{BDJ}{Brans-Dicke-Jordan}
\newacronym{\bbo}{BBO}{Big Bang Observer}
\newacronym{\decigo}{DECIGO}{Deci-Hertz Gravitational-Wave Observatory}

\def\MPR#1{{\it Moving Puncture Recipe}#1 (MPR#1)\gdef\MPR{MPR}}
\def\ahz#1{apparent horizon#1 (AH#1)\gdef\ahz{AH}}
\def\CM#1{center-of-mass#1 (CM#1)\gdef\CM{CM}}
\def\CLA#1{close-limit approximation#1 (CLA#1)\gdef\CLA{CLA}}
\def\pnw#1{post-Newtonian#1 (PN#1)\gdef\pnw{PN}}
\def\nr#1{numerical relativity#1 (NR#1)\gdef\nr{NR}}
\def\qnm#1{quasi-normal mode#1 (QNM#1)\gdef\qnm{QNM}}
\def\isco#1{innermost stable circular orbit#1 (ISCO#1)\gdef\isco{ISCO}}
\def\eos#1{equation of state#1 (EOS#1)\gdef\eos{EOS}}
\def\tov#1{Tolman-Oppenheimer-Volkoff#1 (TOV#1)\gdef\tov{TOV}}
\def\ns#1{neutron star#1 (NS#1)\gdef\ns{NS}}
\def\bbh#1{binary black hole#1 (BBH#1)\gdef\bbh{BBH}}
\def\bhns#1{black hole -- neutron star#1 (BHNS#1)\gdef\bhns{BHNS}}
\def\nsns#1{neutron star -- neutron star#1 (NSNS#1)\gdef\nsns{NSNS}}
\def\emri#1{extreme mass-ratio inspiral#1 (EMRI#1)\gdef\emri{EMRI}}
\def\emrb#1{extreme mass-ratio binaries#1 (EMRB#1)\gdef\emrb{EMRB}} 
\def\grb#1{gamma-ray burst#1 (GRB#1)\gdef\grb{GRB}}
\def\imbh#1{intermediate mass black hole#1 (IMBH#1)\gdef\imbh{IMBH}}
\def\smbh#1{supermassive black hole#1 (SMBH#1)\gdef\smbh{SMBH}}
\def\bh#1{black hole#1 (BH#1)\gdef\bh{BH}}
\def\ulx#1{ultra-luminous x-ray source#1 (ULX#1)\gdef\ulx{ULX}}
\def\nps#1{Newman-Penrose#1 (NP#1)\gdef\nps{NP}} 
\def\lmxbs{low-mass x-ray Binaries (LMXBs)\gdef\lmxbs{LMXBs}\gdef\lmxb{LMXB}} 
\def\lmxb{low-mass x-ray Binary (LMXB)\gdef\lmxbs{LMXBs}\gdef\lmxb{LMXB}}

\newcommand\apjl{\ref@jnl{ApJ}}%
\newcommand\mnras{\ref@jnl{MNRAS}}%

\usepackage[utf8]{inputenc}
\usepackage{multirow}
\usepackage{subfigure}

\usepackage{color}
\usepackage{ulem}   

\providecommand{\e}[1]{\ensuremath{\times 10^{#1}}}

\newacronym{\nr}{NR}{numerical relativity}
\newacronym{\snr}{SNR}{signal-to-noise ratio}
\def\bbh#1{binary black hole#1
  (BBH#1)\gdef\bbh{BBH}}
\def\madm{M^{\mathrm{ADM}}\gdef\madm{M^{\mathrm{ADM}}}}
\def\erad{E^{\mathrm{rad}}\gdef\erad{E^{\mathrm{rad}}}}
\def\jrad{J^\mathrm{rad}}\gdef\jrad{J^\mathrm{rad}}
\def\jinit{J^\mathrm{init}}\gdef\jinit{J^\mathrm{init}}
\def\jzrad{J^\mathrm{rad}_z}\gdef\jzrad{J^\mathrm{rad}_z}
\def\jzinit{J^\mathrm{init}_z}\gdef\jzinit{J^\mathrm{init}_z}
\def\jxrad{J^\mathrm{rad}_x}\gdef\jxrad{J^\mathrm{rad}_x}

\begin{document}

\title{Exploring the Use of Numerical Relativity Waveforms in Burst Analysis of 
Precessing Black Hole Mergers}

\author{Sebastian Fischetti}
\affiliation{Department of Physics, University of Massachusetts, Amherst MA 01003-9337}
\author{Satyanarayan R.P. Mohapatra}
\affiliation{Department of Physics, University of Massachusetts, Amherst MA 01003-9337}
\author{Laura Cadonati}
\affiliation{Department of Physics, University of Massachusetts, Amherst MA 01003-9337}
\author{James Healy}
\affiliation{Center for Relativistic Astrophysics, 837 State Street, Georgia
Tech, Atlanta, GA 30332-0430}
\author{Lionel London}
\affiliation{Center for Relativistic Astrophysics, 837 State Street, Georgia
Tech, Atlanta, GA 30332-0430}
\author{Deirdre Shoemaker}
\affiliation{Center for Relativistic Astrophysics, 837 State Street, Georgia
Tech, Atlanta, GA 30332-0430}

\begin{abstract}
Recent years have witnessed tremendous progress in numerical relativity and an 
ever improving performance of ground-based interferometric gravitational wave 
detectors.  
In preparation for Advanced LIGO and a new era in gravitational wave astronomy, 
the numerical relativity and gravitational wave data analysis communities are 
collaborating to ascertain the most useful role for numerical relativity 
waveforms in the detection and characterization of binary black hole 
coalescences. 
In this paper, we explore the detectability of equal mass, merging black hole 
binaries with precessing spins and total mass  $M_T \in  [80,350]M_{\odot}$, 
using numerical relativity waveforms and template-less search algorithms 
designed for gravitational wave bursts.   
In particular, we present a systematic study using waveforms produced by the 
\mayakranc{} code that are added to colored, Gaussian noise and analyzed with 
the Omega burst search algorithm.  Detection efficiency is weighed against the 
orientation of one of the black-hole's spin axes. 
We find a strong correlation between the detection efficiency and the radiated 
energy and angular momentum, and that the inclusion of the $\ell=2$, 
$m=\pm 1,0$ modes, at a minimum, is necessary to account for the full 
dynamics of precessing systems. 
\end{abstract}
\maketitle


\section{Introduction}
\label{sec:introduction}

The detection of gravitational waves is one of the most exciting developments 
expected for the next decade in Physics and Astronomy.  Ground-based detectors, 
such as LIGO~\cite{Abbott:2007kva} and Virgo~\cite{Acernese2006}, have achieved
their initial design sensitivities and are making progress toward their 
advanced configurations~\cite{Smith:2009bx, aVirgo}. 
According  to stellar population synthesis models~\cite{ratepaper}, Advanced 
LIGO and Advanced Virgo will be sensitive to measurable rates of compact 
binary coalescences, and open a new field of gravitational-wave astronomy. 

There are many uncertainties in the amount and form of gravitational wave 
signals that will be first detected. 
We refer to gravitational wave bursts as transient signatures lasting between 
a few milliseconds and  seconds in the detector sensitive band, as could be
produced by core-collapse supernovae or the late stages of the merger of two
intermediate-mass black holes, as well as serendipitous sources. 
Burst signatures are typically searched for with template-less algorithms 
that identify coincident excess power in multiple 
detectors~\cite{Abadie:2010mt,Abbott:2009zi}.  
In particular, this paper focuses on the detectability of the mergers of 
compact objects with total mass  $M_T \in  [80,350]M_{\odot}$, whose solutions
require simulations of the non-linear Einstein field equations of General 
Relativity.  
The field of \nr{}  has produced solutions to the merger of two black
holes~\cite{Pretorius:2005gq,Campanelli:2005dd,Baker:2005vv,Herrmann:2006ks,
Baker:2006vn,Gonzalez:2006md,Herrmann:2007ac,Koppitz:2007ev, Campanelli:2007ew,
Gonzalez:2007hi, Tichy:2007hk, Campanelli:2007cga, Baker:2007gi,Herrmann:2007ex,
Brugmann:2007zj, Schnittman:2007ij, Pollney:2007ss,Lousto:2007db, Baker:2008md,
Dain:2008ck, Healy:2008js,Gonzalez:2008bi} that are now being used as signals 
to test the performance of analysis routines~\cite{Aylott:2009ya} and to
calibrate the analytical and phenomenological waveform families used as 
template banks in searches~\cite{Buonanno:2007pf,Ajith:2009bn,Sturani:2010yv}.

For systems with total mass greater than $\sim80M_{\odot}$, the final stages 
of a \bbh{} coalescence, merger and ringdown, yield most \snr{} in ground-based
detectors.  
This is the motivation for the work presented in this paper: a systematic study
of how the parameters of the coalescence affect the detectability of mergers of
\bbh{s} in a burst search.  
We test the detectability of mergers in simulated Gaussian noise at the initial
LIGO design sensitivity using the Omega algorithm~\cite{omega}, used by the
LIGO-Virgo collaboration to search for gravitational wave bursts, with a fixed,
single-detector threshold of $\mathrm{SNR} \ge 5.5$,  chosen to match the
definition of single-detector ``trigger'' in compact binary coalescence searches
in real LIGO noise~\cite{Abbott:2009qj}.
We are interested in the effects of both numerical and physical parameters; in
this study, we focus on the effect of spin and its orientation as well as the
number of included harmonics and gravitational-wave cycles.

This paper is organized as follows.  
In Section~\ref{sec:background}, we explain how the waveforms are generated and
how the Omega pipeline burst search algorithm works.  
In Section~\ref{sec:methods}, we explain how we use the \nr{} waveforms to
generate the simulated signals, and discuss the analysis and postprocessing of
the data.  
Results are presented in Section~\ref{sec:results}, and conclusions in
Section~\ref{sec:conclusions}.


\section{Background}
\label{sec:background}

\subsection{\nr{} Waveforms}
\label{subsec:waveforms}

\begin{figure}[tb]
\centering
\includegraphics[width=60mm]{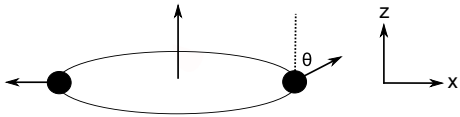}
\caption{Spin configuration used in the simulations for this work.  
The masses and spin magnitudes of the black holes are equal, and the orbit is
initially quasi-circular.}
\label{fig:blackholespin}
\end{figure}

The waveforms used in this work were generated using the vacuum \bbh{} numerical
relativity code \mayakranc{}, also used in previous  
studies~\cite{Healy:2009zm,Healy:2009ir,Herrmann:2007ex,Hinder:2007qu}.
\mayakranc{} adopts the Baumgarte-Shapiro-Shibata-Nakamara formulation with
moving puncture gauge conditions ~\cite{Campanelli:2005dd,Baker:2005vv} and the
\texttt{KRANC} code generator \cite{Husa:2004ip}.  
Bowen-York extrinsic curvature is used to solve the momentum constraint
initially~\cite{Bowen:1980yu}, and the Hamiltonian constraint is solved using
the \texttt{TWOPUNCTURES} spectral solver \cite{Ansorg:2004ds}.  
This study uses simulations of equal-mass black holes initially located on the
$x$-axis, with momentum determined by the 3rd-order post-Newtonian angular
momentum~\cite{Kidder:1995zr,Damour:1999cr}, according to the geometry in
Figure~\ref{fig:blackholespin}.  
Most simulations have an initial separation of $d/M = 6.2$, although two sets 
of simulations with larger initial separations of $d/M = 8$ and $d/M = 10$ 
are also included.   
Our code units are given by $M=m_1+m_2=2m$, where $m$ is the horizon mass of 
the initial black holes.
In all simulations, the black holes are spinning, with $S_- = \{-a,0,0\}m$ and
$S_+ = \{a \sin\theta, 0, a \cos\theta\}m$.  
We vary $\theta$ from $0^\circ$ to $360^\circ$ for a given $a/m$, and we
consider three spin values: $a/m = \{0.2, 0.4, 0.6\}$.   We hold $d/M$, $a/m$
and the total initial mass of the back holes constant as we vary $\theta$.

The grid structure for each run consists of ten levels of refinement provided 
by \texttt{CARPET} \cite{Schnetter-etal-03b}, a mesh refinement package for
\texttt{CACTUS} \cite{cactus-web}.  
We use sixth-order spatial finite differencing; the outer boundaries are located
at $317M$, the finest resolution is $M/77$ and the waveforms are extracted at
$75M$.  
A few waveforms have been generated at resolutions of $\{M/64, M/77, M/90\}$,
with convergence consistent with our sixth-order code.  
The short (long) runs exhibit a phase error on the order of $5\e{-3}$ 
($5\e{-2}$) radians and an amplitude error of $\approx 2\%$ ($\approx 5\%$).
Similar accuracy can be expected in all runs we performed.

\begin{figure}[tb]
\centering
\includegraphics[width=80mm]{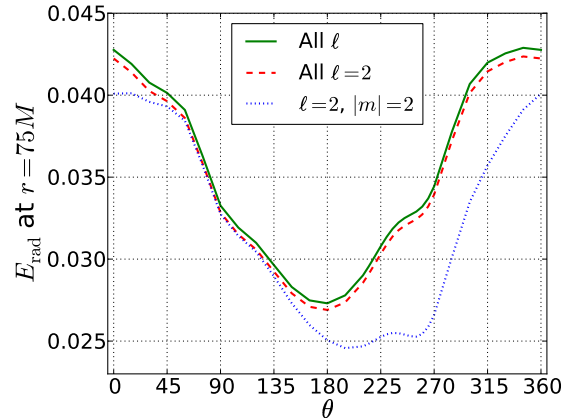}
\caption{Radiated energy, normalized by $\madm$, plotted versus $\theta$ for 
the series with $d/M = 6.2$ and $a/m = 0.6$.  
The radiated energy is computed at an extraction radius of $r=75M$.  
The solid line is all $\ell\le 6$, dashed: $\ell=2$, dotted: dominant 
$\ell=2$, $|m|=2$.}
\label{fig:Erad}
\end{figure}

To characterize the \nr{} runs, in Figure~\ref{fig:Erad} we plot as a function
of $\theta$ the ratio $\erad/\madm$, that is the total radiated energy ($\erad$)
normalized by the total mass of the system given by the Arnowitt-Deser-Misner
mass ($\madm$).   
The radiated energy varies from approximately 2.8\% to almost 4.4 \% of the
total ADM mass.  
The minimum occurs when $\theta \approx 180^\circ$ due to the way the system 
is configured initially.  
By keeping the initial separation fixed, the misalignment of the spin with the
orbital angular momentum results in $\theta=180^\circ$ being the shortest run
with very little inspiral before merger.  
The dashed line was computed using only the dominate mode, $\ell=2$, $m=\pm 2$,
the solid has all modes up to and including $\ell=6$ and the dashed all the
$m$-modes of $\ell=2$  ($m=0$,~$m=\pm 1$,~$m=\pm 2$). Not including all the
$\ell=2$ modes in computing $\erad$ clearly underestimates the energy radiated
for some values of $\theta$.  Modes other than $\ell=2$, however, contribute
little to the total radiated energy, comparing solid and dashed.

\begin{figure}[t]
\centering
\includegraphics[width=80mm]{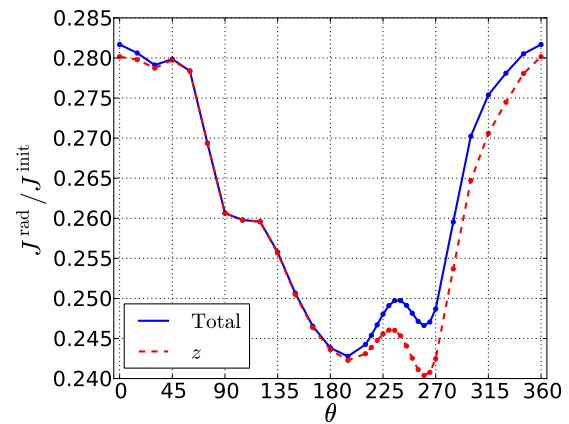}
\caption{The solid curve is the total magnitude of the radiated angular
momentum normalized by the initial angular momentum for each run in the series 
with $d/M = 6.2$ and $a/m = 0.6$. 
The dashed curve is the radiated angular momentum in only the $z$-direction
with the same normalization.}
\label{fig:jrad}
\end{figure}

In Figure ~\ref{fig:jrad}, the solid curve is the total magnitude of the
radiated angular momentum $\jrad$, normalized by the initial angular momentum
$\jinit$, while the dashed curve is the radiated angular momentum in the
$z$-direction $\jzrad$, normalized by $\jzinit$ (the initial angular momentum 
was not fixed for all simulation runs).
When the two curves align, the simulations are dominated by radiation in the
$z$-direction.   
As we approach $\theta=250^\circ$, the other components of the angular momentum
contribute significantly.   
The importance of the numeric value of $\theta$ is only in its labeling of the
binary dynamics, for example that $\jxrad$ is at a maximum for
$\theta=250^\circ$. 

\subsection{Burst Searches}
\label{subsec:burst}

Unmodeled burst searches look for gravitational wave transients in LIGO and 
Virgo data without assumption on the waveform 
morphology~\cite{Abadie:2010mt,Abbott:2009zi}.  
This is accomplished by searching for statistically significant excess signal
energy in the strain time series.  
The advantage of this type of analysis is that, because it does not require
knowledge of the waveform \textit{a priori}, it has the potential to detect
signals missed by other searches. 
The particular burst algorithm used  for this study is the {\it Omega
pipeline}~\cite{omega}.  
This algorithm first whitens the data, then decomposes it into a bank of
windowed complex exponentials characterized by a central time $\tau$, 
frequency $\phi$, and quality factor $Q$.  The signal is tiled in the
$\tau$-$\phi$-$Q$ space, and any tiles with a normalized energy above some
threshold are recorded as possible events.  
This search is equivalent to a matched filtering against whitened data using a 
template bank of sine-Gaussians. 


\section{Methods}
\label{sec:methods}

\begin{figure}[t!]
  \centering
\includegraphics[width=0.45\textwidth]{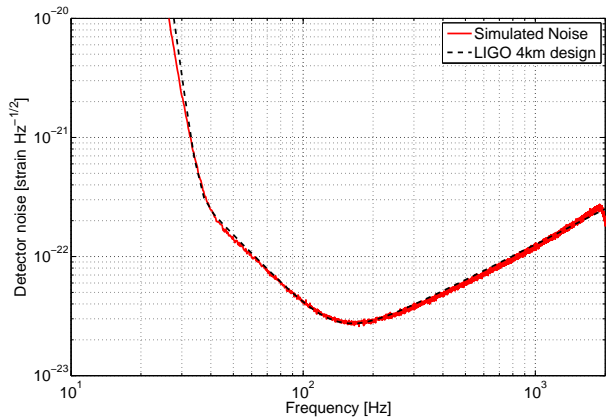}
\caption{The dashed line indicates the design sensitivity of LIGO's 4km
detectors~\cite{Abbott:2007kva}; the continuous line is the spectrum of
simulated noise used in our study. }
\label{fig:spectra}
\end{figure}

The NR waveforms were exchanged according to the data format described
in~\cite{Brown:2007jx}; the transformation between $\Psi_4$ and strain was done
with the method outlined in~\cite{Reisswig:2010di}.
In our study, \nr{} waveforms are used to simulate signals from sources with
random sky location and source inclination; the same set of simulation is scaled
so as to effectively produce signals from~$27$ concentric shells at different
radial distances from the detector. 
These simulated waveform are then injected into Gaussian noise colored to mimic
the sensitivity of the initial LIGO design,  shown in Figure~\ref{fig:spectra}.
The combined signal and noise are then fed into Omega as a simulated detector
output; the results are postprocessed to obtain the detection reach  statistic
described below.

\begin{figure}[t]
\centering
\includegraphics[width=0.45\textwidth]{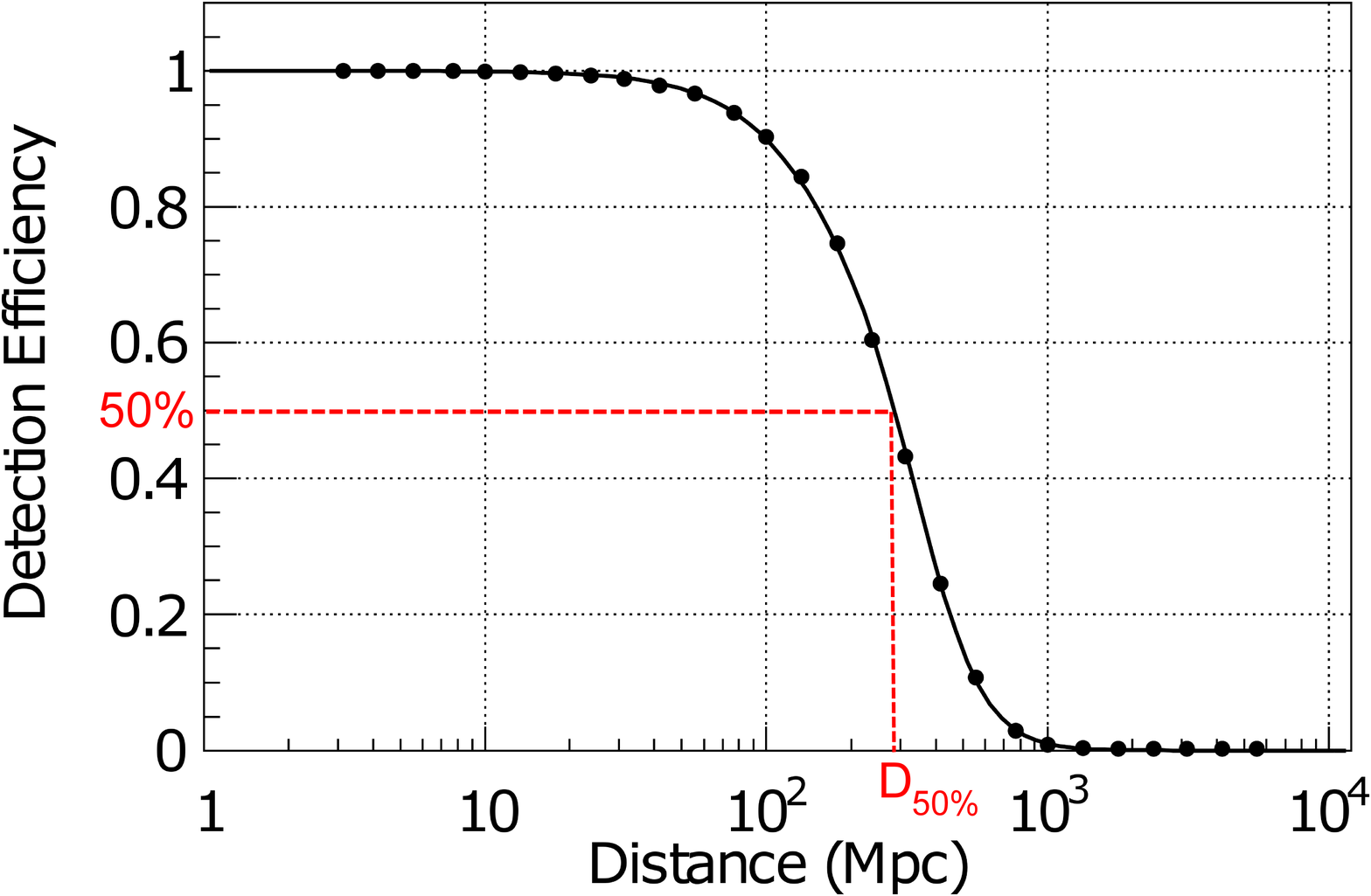}
\caption{A sample plot of detection efficiency as a function of distance to 
the source.  
Each point corresponds to the average detection efficiency of a spherical shell
of radius equal to the distance in abscissa. 
The curve is a sigmoidal fit to the data, used to interpolate the
distance~$D_\mathrm{50\%}$ at which the detection efficiency falls to~$50\%$,
referred to as the {\it reach} of the search. Note that there is an uncertainty
in the value of each point, though the error bars are too small to be visible 
in this plot.}
\label{fig:efficiency}
\end{figure}

Since this study is meant to provide a relatively broad insight into how
efficiently burst searches detect \bbh{} coalescences, we allowed some
simplifications.  
First, we initially used only the dominant quadrupole modes of the waveforms,
$(\ell,m) = (2,\pm 2)$, and disregarded all others.  
We found this has a noticeable impact on the waveform detectability; we discuss
later the effect of using all~$\ell = 2$ modes.
Second, to disentangle waveform effects from that of non-stationary noise in
real interferometer data, we didn't  reproduce a full coherent multi-detector
analysis at fixed false alarm rate, but instead only analyzed data from a
single, ideal detector, chosen to be one of the LIGO 4km detectors, with a 
fixed SNR detection threshold.

This study is restricted to equal-mass systems with a total mass in the
$80-350$~$M_\odot$ range.  Because the frequency evolution of the waveforms
scales as the total mass of the system, this range is chosen so that, in the
initial detectors, most of the \snr{}  comes from the merger and ringdown
portions of the waveform.
For systems with lower total mass, there are enough inspiral cycles in the
sensitive band to make it appropriate to use matched filtering techniques, as
opposed to burst search algorithms~\cite{Aylott:2009ya}.  
For larger values of the total mass, the ringdown frequency falls below the
40~Hz high-pass filter applied to the data, and the waveform falls below the
LIGO seismic noise barrier.  
Our best sensitivity is for systems with total mass in the $150-200$~$M_\odot$
range.

\begin{figure*}[tb]
\centering
\includegraphics[width=133mm]{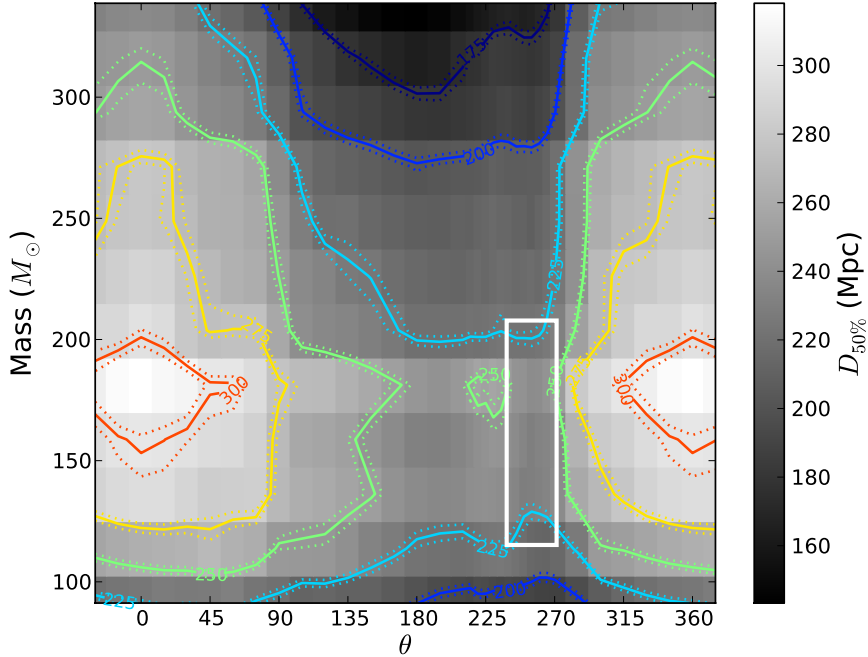}
\caption{Reach for the system discussed in Section~\ref{sec:results}, where 
the adimensional spin of each black hole is~$0.6$, using only the
$(\ell, m) = (2, \pm 2)$ modes of the waveforms.  
To clarify features at the left and right edges of the plot, four angles were
repeated.  
Dotted lines in the contours represent uncertainties from the limited statistics
of the runs.  
Notice that the trend appears to go roughly as the total angular momentum of the
system: when $\theta=0^\circ$, one of the black holes is spin ``up'',  parallel
to the orbital angular momentum; when~$\theta=180^\circ$, it is spin ``down'',
antiparallel to the orbital angular momentum.
This trend is broken by an asymmetry near~$\theta=255^\circ$ for masses
between~$125$ and~$300$~$M_\odot$; the white rectangle marks the location 
where this is most pronounced.}
\label{fig:Mvstheta}
\end{figure*}

Once the Omega algorithms identifies instances of excess power in the simulated
data, we compare the trigger list  to the list of injections to identify which
were missed and found.  
We use an \snr{} threshold of 5.5, chosen to match the  definition of
single-detector ``trigger'' in compact binary coalescence searches in real LIGO
noise~\cite{Abbott:2009qj}.   
This threshold choice is one of convenience, for a study of the dependence of
detectability on various other simulation parameters. 
In actual LIGO searches, the detection threshold will be determined by the rate
of non-Gaussian noise transients and what is considered to be an acceptable
false alarm probability~\cite{Abadie:2010mt,Abbott:2009zi}.
For each concentric shell in our simulated sky, we compute an average detection
efficiency, plot it as a function of the shell radius and fit a sigmoid to the
data; to quantify the detection efficiency of that search, we use this fit to
interpolate the distance at which the detection efficiency is~$50$\%, herein
referred to as {\it reach}  and denoted by~$D_\mathrm{50\%}$.  
In Figure~\ref{fig:efficiency}, we show the particular fit for the case
$a/m=0.6$, $\theta=0^\circ$, $d/M=6.2$.


\section{Results}
\label{sec:results}
This study focuses on the dependence of detection efficiency on the orientation
and magnitude of the spins of the two merging black holes.  

The system under study consists of two equal-mass black holes with the same 
spin magnitude.  
In the coordinate system where the initial orbital angular momentum of the
system is in the $+z$ direction, the initial spin of one black hole is in
the $-x$ direction, while the initial spin of the second black hole is 
oriented at an angle~$\theta$ from the~$+z$ direction, as sketched in
Figure~\ref{fig:blackholespin}.  
We examine the dependence of detection efficiency on both the magnitude of 
the spins and the parameter~$\theta$.

\subsection{Dependence on total mass}
We begin with a system where the magnitude of the adimensional spin of each
black hole is~$0.6$.  
Numerical simulations were generated for $\theta$ ranging from~$0^\circ$ to
$345^\circ$, and the mass was binned into twelve bins, $22.5$~$M_\odot$ wide, 
to produce a contour plot of the reach in the~$M-\theta$ plane  as shown in
Figure~\ref{fig:Mvstheta}.  
For a given value of~$\theta$,  the reach peaks at~$\approx 180$~$M_\odot$. 
This reflects how the waveform frequency structure scales with the system total
mass: at~$\approx 180$~$M_\odot$, the merger falls into the detector's most
sensitive frequency band.  
As the total mass of the system either increases or decreases, the waveform
portion triggering Omega correspondingly moves later or earlier into the
waveform, where less power is emitted, and consequently the reach is
reduced~\cite{Aylott:2009ya,2009CQGra..26t4005C}.  
This feature is well understood, and not unique to this particular system.

\subsection{Dependence on spin orientation}
Figure~\ref{fig:Mvstheta} shows that, at a given mass, the reach peaks at
$\theta=0^\circ$ and has a minimum near~$\theta=180^\circ$, though its precise
location depends on the total mass.  
Keeping in mind that at~$\theta=0^\circ$ one of the black holes is spin ``up'',
parallel to the orbital angular momentum, and at~$\theta=180^\circ$ it is spin
``down'', antiparallel to the orbital angular momentum, the reach appears to go
roughly as the radiated energy and angular momentum of the system.  
The reach deviates from the total energy radiated in the vicinity of
$\theta=255^\circ$ and is instead more evocative of  $\jzrad$ and of the
dominant $(\ell, m)=(2, \pm 2)$ modes contribution to $\erad$, as seen in
Figures~~\ref{fig:Erad} and~\ref{fig:jrad}, respectively. 

\begin{figure}[tb]
\centering
\includegraphics[width=80mm]{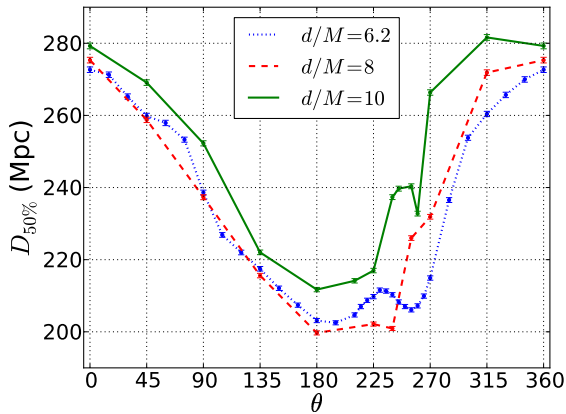}
\caption{Average reach in $80-350$~$M_\odot$ as a function of $\theta$ for
systems with different initial black hole coordinate separations.  
The dotted  line is~$d/M = 6.2$, dashed is~$d/M = 8$, and solid is~$d/M = 10$.}
\label{fig:NRlength}
\end{figure}

\begin{figure}[th]
\centering
\subfigure[]{
\includegraphics[width=80mm]{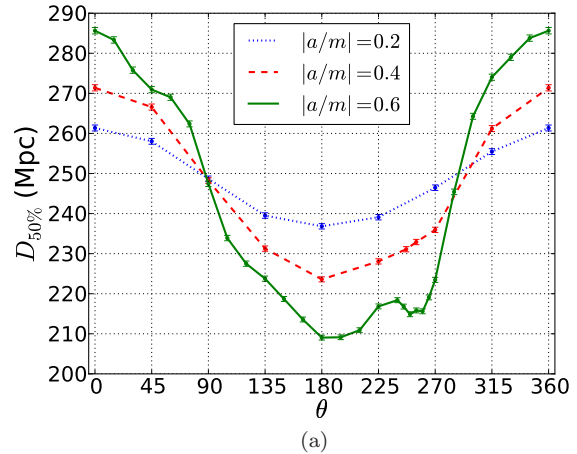}
\label{subfig:spineff}
}
\subfigure[]{
\includegraphics[width=80mm]{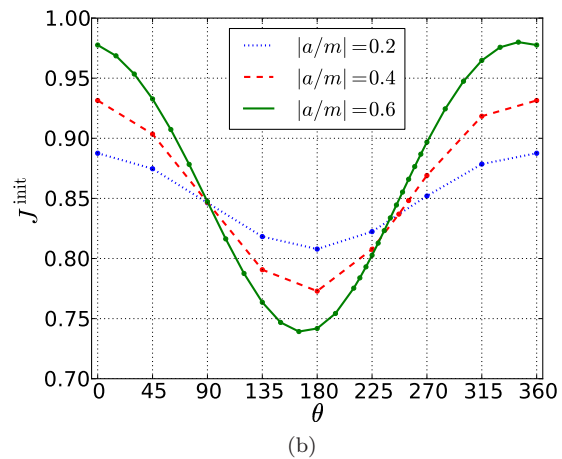}
\label{subfig:jinit}
}
\label{fig:spins}
\caption{Panel \subref{subfig:spineff}:  reach as a function of~$\theta$ for 
the system sketched in Figure~\ref{fig:blackholespin} with $a/m=0.2$,~$0.4$,
and~$0.6$, averaged over $80-350$~$M_\odot$:  $a/m=0.2$ is the dotted  line,
$a/m=0.4$ the dashed, $a/m=0.6$ the solid.  
The three systems exhibit a similar  behavior, proportional to the spin
magnitude. 
Panel \subref{subfig:jinit}: the initial angular momentum is plotted versus
$\theta$ for the same three spin magnitudes.}
\end{figure}

\subsection{Dependence on initial coordinate separation}

Since the waveforms used to generate Figure~\ref{fig:Mvstheta} were relatively
short, with only a few cycles before merger, we repeated the analysis with
waveforms of different lengths and compared simulations with initial coordinate
separations~$d/M$ of~$6.2$,~$8$, and~$10$; the system in
Figure~\ref{fig:Mvstheta} has $d/M=6.2$.  
The results, averaged over the entire mass range, are shown in 
Figure~\ref{fig:NRlength}.  
The individual spins of the black holes, as well as the orbital angular
momentum, will in general precess about the total angular momentum of the
system. 
Therefore, two simulations of different lengths with the same initial value of
$\theta$ will in general \textit{not} represent the same system.  
We are interested, however, not in confirming point-by-point agreement between
the simulations, but rather agreement in the overall behavior of the reach for
the different simulations lengths. 
All three simulation sets exhibit the same sinusoidal behavior observed in the
shorter simulations and, to  varying degree, the asymmetry at
$\theta=255^\circ$.  
The effect is more subtle in the simulations with $d/M = 8$  due to sparse
$\theta$ sampling in the region surrounding the $\theta=255^\circ$.

\begin{figure}[tb]
\centering
\includegraphics[width=80mm]{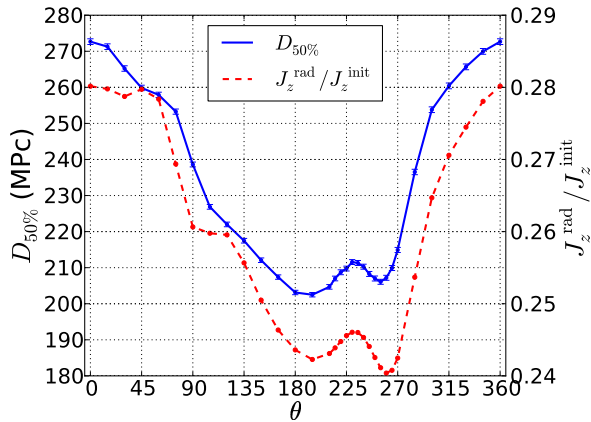}
\caption{Average reach in the $80-350$~$M_\odot$ mass range, as a function of
$\theta$, compared to the normalized~$z$-component of the radiated angular
momentum. Only the dominant $m=\pm2$ modes are included.}
\label{fig:DistJz}
\end{figure}

\begin{figure}[th]
\centering
\includegraphics[width=80mm]{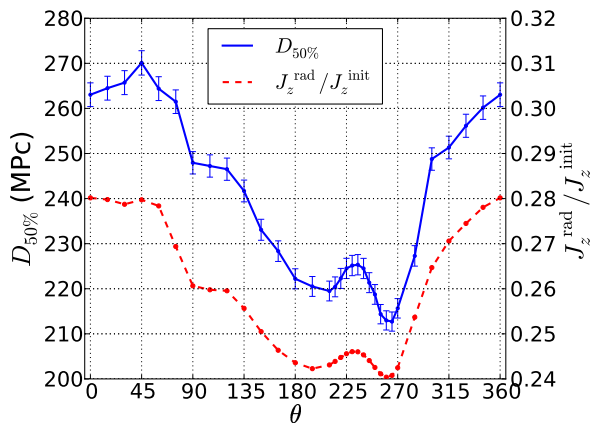}
\caption{Reach as a function of~$\theta$, restricted to $102.5-125.0$~$M_\odot$,
compared to the normalized $z$-component of the radiated angular momentum.  
Only the  dominant $m=\pm2$ modes are included.}
\label{fig:massrange}
\end{figure}

\subsection{Dependence on spin magnitude}
Next, we compare the behavior of the same system for adimensional spin 
magnitudes~$0.2$,~$0.4$, and~$0.6$.   
For a comparative analysis, we forgo breaking up the data into mass bins, and
instead show the average reach in the $80-350$~$M_\odot$ mass range  in
Figure~\ref{subfig:spineff}.    
All three systems exhibit the same bulk behavior, with a peak at
$\theta=0^\circ$ and a minimum at~$\theta=180^\circ$.  
The asymmetry in $\theta$ is evident for spin~$0.6$, but it does not show in 
the lower spin cases in Figure~\ref{subfig:spineff}, due to the combination of 
two factors: for practical reasons, $\theta$ was sampled less at spin 0.4 and 
0.2 than at spin 0.6, and the amount of angular momentum radiated in the
$x$-direction decreases with decreasing spin, as the precession is decreasing.
Overall, we find that the variation of the reach is proportional to the
magnitude of the black hole spins, which is consistent with the observation 
that the reach scales with the total angular momentum of the system.  
For comparison, Figure~\ref{subfig:jinit} shows how the  initial angular
momentum of the system depends on the spin magnitude, as a function of the
angle~$\theta$.

\begin{figure}[tb]
\centering
\includegraphics[width=80mm]{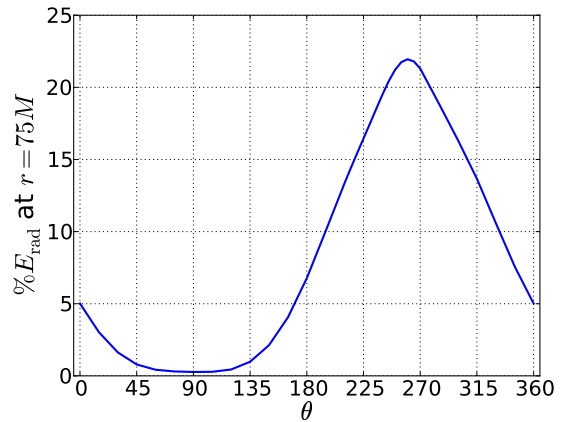}
\caption{Percentage of energy radiated by the non-dominant modes as a 
function of $\theta$.}
\label{fig:Epercentrad}
\end{figure}

\subsection{Correlation with radiated energy and angular momentum}

The asymmetry of the reach versus $\theta$, seen near $\theta=255^\circ$ is
over-emphasized, since our original analysis did not include all the $\ell=2$ 
modes.  
Figures~\ref{fig:DistJz}  and \ref{fig:massrange} show the reach calculated 
with only the $\ell=|m|=2$ modes versus $\theta$ for two choices of total 
mass for the system with $a/m=0.6$ and $d/M=6.2$.  
In Figure~\ref{fig:DistJz}, the reach is averaged over the entire mass range 
of $80 - 350 M_\odot$ while in Figure~\ref{fig:massrange} one bin of total 
mass is selected: $102.5-125.0 M_\odot$. 
In each of these plots,  $\jzrad/\jzinit$ is included for reference.    

\begin{figure*}[th]
\centering
\includegraphics[width=133mm]{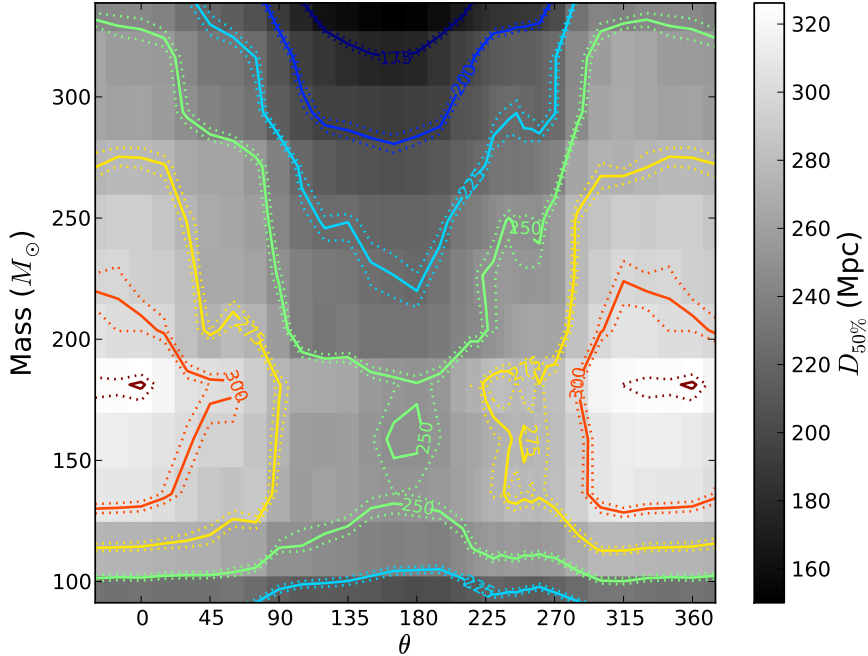}
\caption{Same as Figure~\ref{fig:Mvstheta}, but using all~$\ell = 2$ modes 
of the waveforms.  
Notice that the asymmetry at~$\theta = 255^\circ$ is significantly reduced by 
the addition of the additional modes.}
\label{fig:allmodes}
\end{figure*}

\begin{figure}[tb]
\centering
\includegraphics[width=80mm]{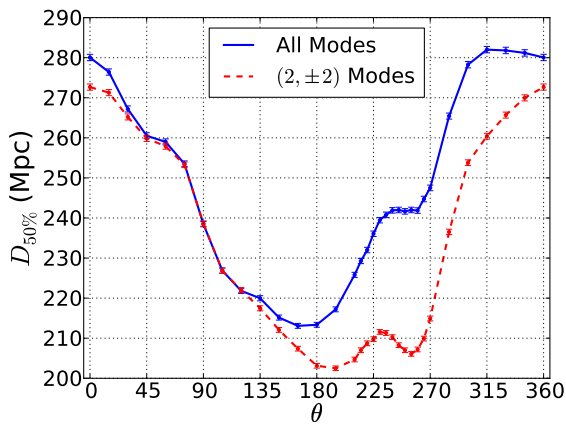}
\caption{Reach as a function of~$\theta$ averaged over the entire mass
range~$80-350$~$M_\odot$ for injections using only the dominant
$(\ell, m) = (2,\pm 2)$ modes and injections using all~$\ell = 2$ modes.  
Notice that in the vicinity of~$\theta = 90^\circ$, the curves agree very 
well, because this is where the black hole spins are antiparallel and cancel,
making the contributions of the~$m = 0$ and~$m = \pm 1$ modes negligible.
As~$\theta$ moves away from~$90$, the additional modes become more important,
and the curves differ very significantly.}
\label{fig:modecomp}
\end{figure}

Comparing the reach calculated from just the $\ell=|m|=2$ modes with 
that calculated from all the
$\ell=2$ modes in Figures~\ref{fig:allmodes} and \ref{fig:modecomp}, it is 
clear that the reach asymmetry is alleviated with the inclusion of all the
$\ell=2$ modes. 

The dependence of the reach on $z$-component $\jzrad$, rather than the total
radiated angular momentum $\jrad$, when only the  $\ell=2$, $m=\pm 2$ modes 
were used in the calculations, is explained by Equations (3.22)-(3.24)
from~\cite{2008GReGr..40.1705R}, where the radiated momentum in the $x$ and
$y-$directions depends on non-zero $m\neq \ell$ harmonics.   
Likewise, we get the same effect if one ignores the $m\neq \ell$ modes in
calculating the radiated energy.  
Figure~\ref{fig:Epercentrad} gives a clear indication of why the value 
$\theta=255^\circ$ appears ``special'' in our choice of initial data by 
plotting the percent energy radiated in the non-dominant $\ell=2$ modes.  

\begin{figure}[tb]
\centering
\includegraphics[width=80mm]{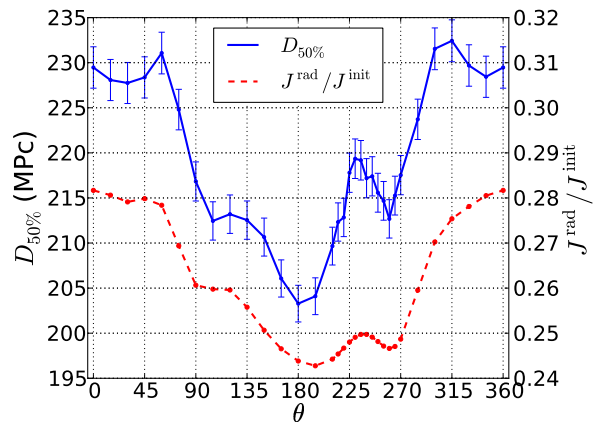}
\caption{The solid  curve is the reach computed using all~$\ell = 2$ modes,
constricted to the lowest mass range~$80-102.5$~$M_\odot$.  
This is compared to the radiated angular momentum~$\jrad$ normalized to the
initial angular momentum~$\jinit$, the red dashed curve.  Note that the two
curves appear to have very similar behavior, indicating that the detection
efficiency of injection including all~$\ell = 2$ modes is roughly proportional
to the radiated angular momentum.}
\label{fig:allmodeslowmass}
\end{figure}

The reach calculated with all the $\ell=2$ modes and averaged over the 
$80-350 M_{\odot}$ mass range, shown in Figure~\ref{fig:modecomp}, is well
represented by the radiated energy.  
The reach, however, does depend on the mass range, as seen for instance in
Figure~\ref{fig:allmodeslowmass}.  
In particular, the lower mass bins of data are well described by the radiated
angular momentum rather than the radiated energy.   
We speculate that  at low masses, the Omega algorithm tends to trigger earlier
in the waveform, when most of the angular momentum is being radiated. 
On the other hand, when averaged over the full mass-scale, the reach takes on
the characteristic of radiated energy, shown in Figure~\ref{fig:Erad} as we
would have expected.

\section{Conclusions}
\label{sec:conclusions} 
We used simulated waveforms for \bbh{} coalescences, calculated with the
\mayakranc{} code, to characterize the detection efficiency of the LIGO-Virgo
Omega burst algorithm as a function of the morphological parameters of the
system and to investigate the algorithm's response  to bulk features in
gravitational waves from \bbh{} mergers.  
We chose one particular set of equal-mass, quasi-circular, spinning waveforms
that allow for precession; for fixed initial spin magnitudes and fixed initial
separation of the black holes in the simulations, we  investigated the
dependence of the detection efficiency on the orientation of the initial spin
vector of one of the black holes, parametrized by the angle $\theta$ between
such spin and the angular momentum of the system.

As $\theta$ is varied, the initial angular momentum changes as in
Figure~\ref{subfig:jinit}, resulting in different dynamics of the binary.  
The detection reach, established with the Omega burst algorithm, is sensitive 
to such differences, peaking when the initial  black-hole spin vector is
$\theta=0^\circ$ and dipping near  $180^\circ$, consistent with the system's
initial angular momentum.
A departure from this trend, noted near $\theta=255^\circ$, is an indication
that non-dominant modes are important in these systems.
If all the~$\ell = 2$ modes are included in the simulations, the reach is well
understood by the radiated energy when averaging over the entire mass range
studied,  $80-350M_\odot$; however, when the reach is computed at the low-end of
the mass range its behavior is more evocative of radiated angular momentum.   
We conclude that it is important to account for all $m\ne\ell$
modes to both compute the full radiated angular momentum  and to measure the
detection reach of a burst search for precessing \bbh{} systems.

For non-precessing, aligned-spin, equal-mass systems, the higher the magnitude
of the black-hole spins, the greater the reach, since the system radiates more
angular momentum~\cite{2009PhRvD..80l4026R}. 
For precessing systems, we find the reach depends on the orientation of the
spins: higher-spin systems have a greater reach when the $\jzinit$ is at a
maximum and the spin is aligned with the orbital angular momentum, near
$\theta=0^\circ$, but a lower reach when $\jzinit$ is at a minimum, when the
spin anti-aligned, near~$\theta=180^\circ$. 
We find the variation of reach with~$\theta$ is proportional to the magnitude of
the black hole spins. 
For a system with two equal magnitude spins of~0.6, the reach variation with
orientation is~$\sim 30\%$, resulting in a~$\sim 120\%$ difference in sensitive
volume and in the rate of detectable precessing \bbh{} in initial and advanced
interferometric detectors.

The parameter space of \bbh{s} is large and this work targets only a  portion 
of it.  
In future studies, we will continue to explore the search algorithm's response
to the parameters of precessing systems, such as the black hole mass ratio,
and investigate how degeneracies in the detectable parameter space caused by the
transition from a largely parametrized inspiral phase to a two-parameter end
state impacts burst algorithms' ability to extract the system's physical
properties.

\begin{acknowledgements}
This work is supported by NSF grants to LC PHY-0653550 and PHY-0955773 and  by
NSF grants to DS PHY-0925345, PHY-0941417, PHY-0903973, PHY-0955825 and
TG-PHY060013N. 
We thank Pablo Laguna, Ian Hinder, Frank Herrmann and Tanja Bode for their
contributions to the \mayakranc{} code, as well as Shourov Chatterji, Jameson
Rollins, Antony Searle  and the LIGO Scientific Collaboration for their
contributions to the Omega Burst Search Algorithm. 
We also thank members of the Numerical INJection Analysis (NINJA) collaboration,
for useful discussions on this subject.
This paper is assigned LIGO Document Control Center number P1000092.

\end{acknowledgements}

\bibliography{references,ninja}

\end{document}